# Estimation of residual carrier density near the Dirac point in graphene through quantum capacitance measurement

K. Nagashio*, T. Nishimura, and A. Toriumi
Department of Materials Engineering, The University of Tokyo, Tokyo 113-8656, JAPAN
*E-mail: nagashio@material.t.u-tokyo.ac.jp

## ABSTRACT

We discuss the residual carrier density ($n^*$) near the Dirac point (DP) in graphene estimated by quantum capacitance ($C_Q$) and conductivity ($\sigma$) measurements. The $C_Q$ at the DP has a finite value and is independent of the temperature. A similar behavior is also observed for the conductivity at the DP, because their origin is residual carriers induced externally by charged impurities. The $n^*$ extracted from $C_Q$, however, is often smaller than that from $\sigma$, suggesting that the mobility in the puddle region is lower than that in the linear region. The $C_Q$ measurement should be employed for estimating $n^*$ quantitatively.

The extraction of the quantum capacitance ($C_Q$) through capacitance measurements of graphene provides direct information on the density of states (*DOS*) of graphene because it is regarded as the energy cost of inducing carriers in graphene and is directly related as $C_Q=e^2DOS$. In the field-effect transistor (FET) structure, $C_Q$ is introduced in series with the geometrical capacitance ($C_{ox}$) in the equivalent circuit ($1/C = 1/C_{ox} + 1/C_Q$).[1] The experimentally determined $C_Q$ value near the Dirac point (DP) is, however, larger than that expected from the ideal *DOS* of graphene.[2-6] This indicates that mobile carriers should exist near the DP in graphene FETs. The origin of these residual mobile carriers at the DP is still under the debate. The temperature-independent resistivity at the DP of graphene FETs fabricated on $SiO_2$ substrates suggests that the carriers are induced externally from charged impurities on/in the $SiO_2$ substrates.[7-10] In fact, the temperature-dependent resistivity due to the intrinsic thermal excitation of carriers is observed for suspended graphene[11] and graphene on h-BN substrates[12] as the charged impurity densities ($n_{imp}$) are reduced. These charged impurities not only induce mobile carriers around the DP but also degrade the carrier mobility ($\mu$) in the on-state of graphene FETs; thus, the quantitative estimation of the residual (mobile) carrier density ($n^*$) is quite important. The region around the DP may be understood intuitively by the electron-hole puddle state, which was demonstrated by scanning the single-electron transistor probe.[13] Furthermore, the relationship between $n_{imp}$ and $n^*$ has been predicted theoretically.[9] However, the quantitative comparison between them is limited. In addition, although the dominant scattering mechanism affecting $\mu$ in back-gated graphene FETs is considered to be the Coulomb scattering by charged impurities,[8,9] the scattering mechanism in top-gated graphene FETs is not yet clear.

In this study, dual-gated (top- and back-gated) graphene FETs were fabricated to estimate $n^*$ through $C_Q$ measurements. Then, the dominant scattering source in graphene FETs with the high-*k* top-gate insulator is also discussed from the temperature dependence of the FET mobility.

Monolayer graphene was transferred by the mechanical exfoliation of Kish graphite onto ~90 nm $SiO_2/n^+$-Si substrates (0.01 Ωcm). The $SiO_2$/Si substrates were annealed at 1000 °C for 5 min in an 100% $O_2$ gas flow prior to the graphene transfer process because no hysteresis in drain current ($I_{SD}$) - back-gate voltage ($V_{BG}$) curves has been achieved due to the hydrophobic nature of the siloxane surface of $SiO_2$ substrates.[14] The source and drain electrodes (Ni(~10 nm)/Au(~50 nm)) were deposited by the thermal evaporation after the resist patterning by conventional electron-beam lithography. The devices were annealed in an $H_2$/Ar gas mixture for 1 hr at 300 °C to remove the residual resist. Then, $Y_2O_3$ was deposited over the entire surface area of the wafer by the thermal evaporation of ~2.5 mg of Y metal at $P_{O2} = 10^{-1}$ Pa and at room temperature. Detailed fabrication method is available in supplemental information. Afterward, the wafer underwent the annealing at 200°C for 10 min in an 100% $O_2$ gas flow.[15] The D band was not detected in the Raman measurement through the $Y_2O_3$ layer, which suggests that no noticeable defects were formed during the deposition. The top-gate electrode (Ni/Au) was also patterned by electron-beam lithography. Finally, the device was annealed at 300 °C for 30 s in a 0.1% $O_2$ gas flow before electrical measurements were conducted in a vacuum probe station. An

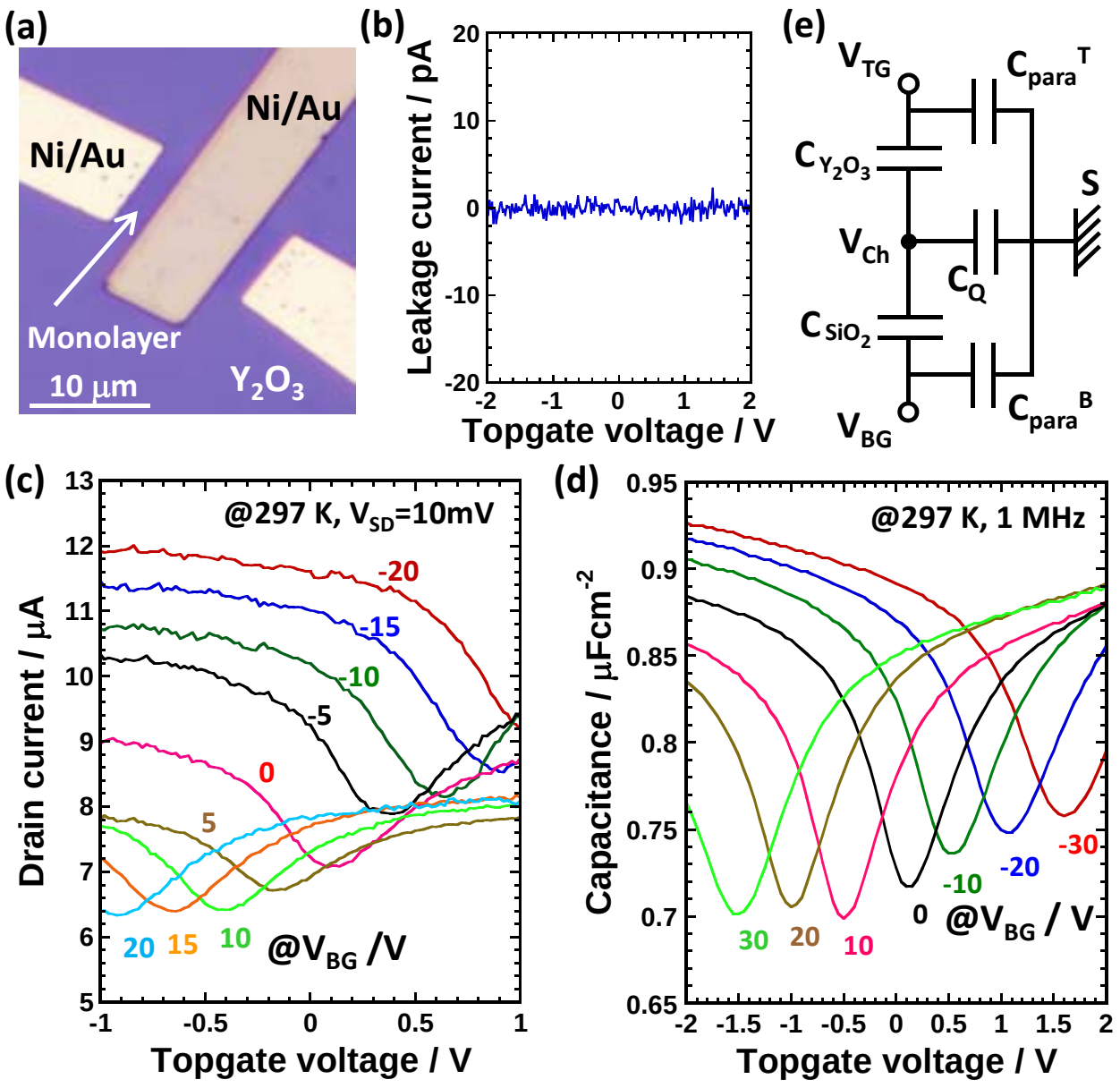

**FIG 1** (color online) (a) An optical micrograph of a dual-gated graphene FET device with an $Y_2O_3$ top-gate insulator. (b) The leakage current as a function of $V_{TG}$ when $I_{SD}$ is 5~10 μA for $V_{BG}$ = 0V at 297 K. (c) $I_{SD}$ as a function of $V_{TG}$ for different $V_{BG}$ at 297 K. (d) $C_{Total}$ as a function of $V_{TG}$ for different $V_{BG}$ at a frequency of 1 MHz at 297 K. (e) The equivalent circuit of the dual-gated device.

optical microscopy image of the typical dual-gated FET is shown in Fig. 1(a). To avoid macroscopically inhomogeneous channel potentials due to the charge transfer from the source and drain contacts,[16] the distance between the top gate and the source (drain) electrodes was designed to be longer than ~2 μm.

Here, we should mention about the reason why $Y_2O_3$ was chosen for the top-gate insulator. The oxidation ability of Y is the highest among high-*k* materials and also higher than that of C based on the standard Gibbs free energy changes for oxidation. Therefore, it is expected that $Y_2O_3$ could be obtained at the relatively low oxidation temperature and is thermodynamically stable on graphene. The electrical characteristics of the thin $Y_2O_3$ films are critical in this study. The leakage current between the source and top-gate electrodes was smaller than a few pA and three orders lower than $I_{SD}$, as shown in Fig. 1(b). This low leakage current was achieved by optimizing $P_{O2}$ during the metallic Y deposition. Figure 1(c) shows $I_{SD}$ as a function of top-gate voltage ($V_{TG}$) for different $V_{BG}$. The top gate controls the graphene channel just below the top-gate electrode, while the back gate (the $n^+$-Si substrate) changes the carrier density ($n$) over the entire area of the graphene channel. Therefore, when the Fermi level ($E_F$) was initially placed deeply in the valence band by $V_{BG}$ = - 20 V and then $V_{TG}$ was swept within ±1 V, $I_{SD}$ was larger than the currents observed for other $V_{BG}$ because the resistance of the access region between the top-gate and source (drain) electrodes was kept low ($p$-type). On the other hand, the total capacitance ($C_{Total}$) between the source and top-gate electrodes for the same device was measured as a function of $V_{TG}$ for different $V_{BG}$, as shown in Fig. 1(d). The equivalent circuit of the dual-gated FET is shown in Fig. 1(e), where $V_{ch}$, $C_{para}^T$ and $C_{para}^B$ are the channel voltage, the parasitic capacitances for the top gate and back gate, respectively. The large dependence of $C_{total}$ on $V_{TG}$ (that is, $E_F$) directly indicates a large contribution of $C_Q$ to $C_{total}$, because $C_{ox}$ is independent of $E_F$.

Contrary to the *I-V* characteristics, the *C-V* characteristic is shifted in parallel with changes in $V_{BG}$ because the capacitance in graphene was modulated just below the top-gate electrode. The slight variation in the minimum $C_{total}$ could be due to the depletion layer formation in the Si substrate (~$5\times10^{18}$ cm$^{-3}$). To confirm this, all *C-V* curves in Fig. 1(d) were superimposed, except for the $V_{BG}$ = ±30 V cases, as shown in Fig. 2(a). It is clear that all of the curves are consistent, suggesting that the parasitic capacitance has no $V_{TG}$ dependence and that the linear dispersion is retained under the external electric field, unlike in the bilayer case.[17]

The hysteresis in *C-V* curves is also a measure of the qualities of both the oxide layer and the graphene/oxide interface. The hysteresis is defined as $\Delta V_{hys} = V_{DP}^{forward} - V_{DP}^{reverse}$ for the $V_{TG}$ sweep, where $V_{DP}$ is the Dirac point voltage. The inset in Fig. 2(a) shows $\Delta V_{hys}$ as a function of temperature for a $V_{TG}$ sweep range of ±2 V. The room temperature data is only shown for $\Delta V_{TG}$ ±1 V. The $\Delta V_{hys}$ is negligible for $V_{TG}$ = ±1 V and - 0.1 V for $V_{TG}$ = ±2 V at room temperature. The fact that $\Delta V_{hys}$ was suppressed by lowering the measurement temperature may suggest the orientation polarization of water molecules.[14,18,19] Therefore, the present $Y_2O_3$ serves quite well as the top-gate insulator.

Systematic shifts in the DP as a function of $V_{BG}$ are evident in the *I-V* and *C-V* measurements, as shown in Figs. 1(c) and (d). The DP voltages obtained for the $V_{TG}$ sweeps in both measurements are plotted as a function of $V_{BG}$ in Fig. 2(b). The DP is controlled by the relative ratio of capacitive couplings between top and back gates with graphene. Therefore, the slope in Fig. 2(b) corresponds to $C_{SiO2}/C_{Y2O3}$.[20] Both lines have the same slope, indicating that both top and back gates capacitively control the dual-gated FET very consistently in *I-V* and *C-V* measurements. Because $C_{SiO2}$ is estimated to be 0.039 μF/cm$^2$ for 88-nm-thick $SiO_2$, $C_{Y2O3}$ can be calculated to be 0.76 μF/cm$^2$. This method is very useful because $C_{Y2O3}$ can be determined without the information on the dielectric constant and thickness of $Y_2O_3$ on graphene. The precise and independent determination of these two quantities for the very limited area of graphene includes large amount of ambiguity.

Next, let us estimate $C_Q$. Figure 3(a) shows the measured capacitance between the source and top gate electrodes, $C_{total}$, as a function of $V_{TG}$ - $V_{DP}$ at $V_{BG}$ = 0 V. The equivalent circuit in Fig. 1(e) can be reduced to that in the inset of Fig. 3(a) due to $V_{BG}$ = 0V. Because $C_{Y2O3}$ is 0.76 μF/cm$^2$ as mentioned above, $C_{para}^T$ is an only fitting parameter to extract $C_Q$. Therefore, the $C_{total}$ value at $V_{TG}$ = ±2 V was adjusted by the fitting of $C_{para}^T$. The dotted lines in Fig. 3(a) are the fitted curves, assuming $C_{para}^T$ = 0.19 μF/cm$^2$.

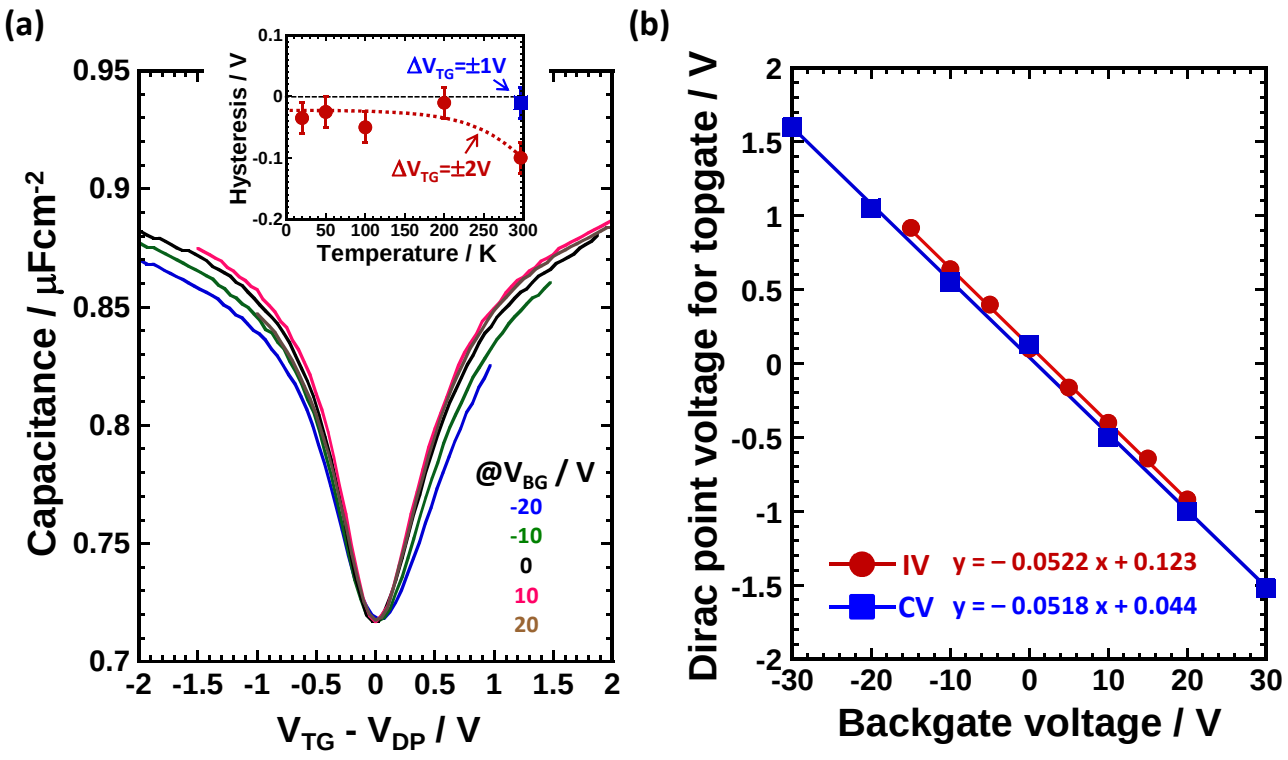

**FIG 2** (color online) (a) $C_{Total}$ as a function of $V_{TG}$ - $V_{DP}$, where the *C-V* curves for $V_{BG}$ = ± 20 V in Fig. 1(d) are superimposed relative to the DP of *C-V* curve at $V_{BG}$ = 0 V for both the transverse and vertical axes. The inset is the temperature dependence of $\Delta V_{hys}$ for the sweeping range of $\Delta V_{TG}$ ±2 V. The room temperature data is only shown for $\Delta V_{TG}$ ±1 V. Both data is obtained for $V_{BG}$=0V. (b) The DP voltage for the $V_{TG}$ sweep as a function of $V_{BG}$. The solid circles and solid boxes indicate the DPs obtained by *I-V* and *C-V* measurements in Figs. 1(c) and (d), respectively. The linear functions obtained by the fitting are shown in the figure.

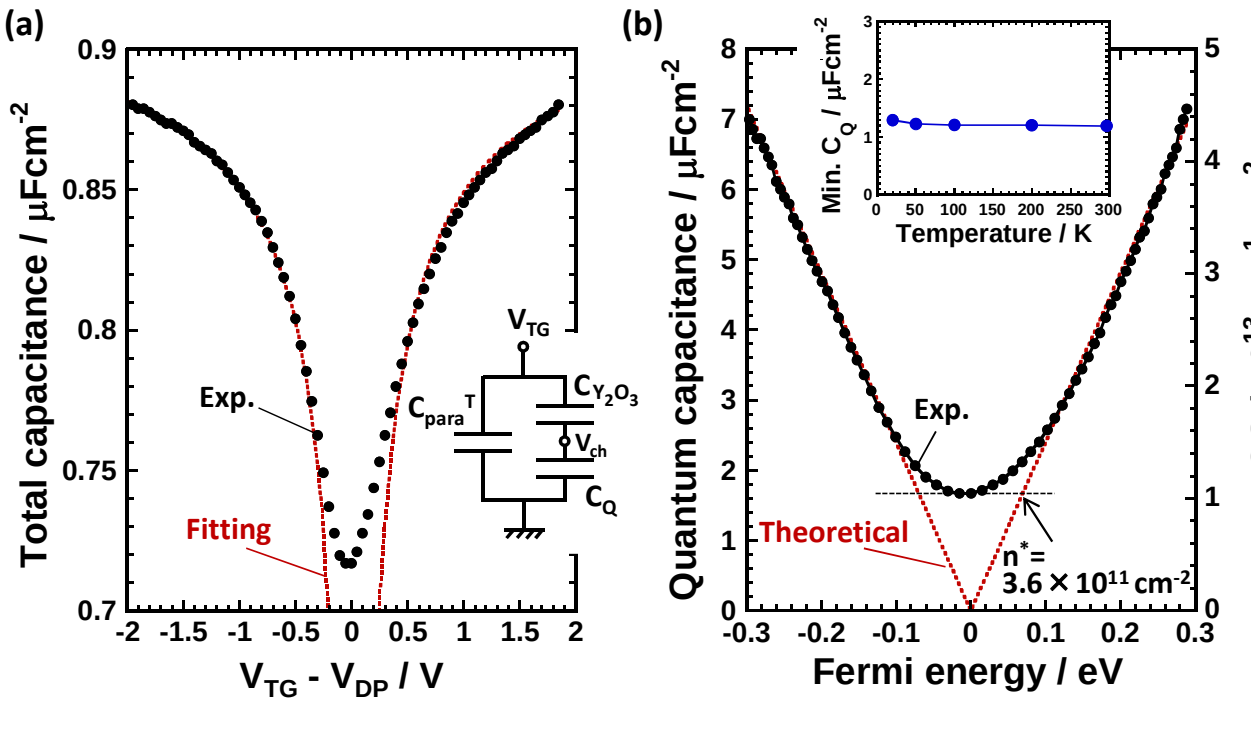

**FIG 3** (color online) (a) $C_{Total}$ as a function of $V_{TG}$ - $V_{DP}$. The solid circles are experimental data; the dotted lines are curves fitted by $C_{para}$ = 0.19 μF/cm$^2$. The inset shows the equivalent circuit for $V_{BG}$ = 0, where $C_{para}$ is in parallel with $C_{ox}$ and $C_Q$. (b) $C_Q$ extracted based on the fitting in (a). The theoretically predicted $C_Q$ is shown by red dotted lines. The right vertical axis indicates the *DOS*. The inset shows the minimum $C_Q$ as a function of temperature for a different sample.

The dominant source for $C_{para}{}^{T}$ could be the capacitance between the topgate and source electrodes. The experimental results are reproduced well, except near the DP. Based on this fitting, $C_Q$ is experimentally estimated as a function of $E_F$, as shown in Fig. 3(b). $E_F$ is indeed the charging energy and is expressed as $E_F=eV_{ch}$. When the serial capacitance is described as $1/C' = 1/C_{Y2O3} + 1/C_Q$ and $V_{TG}'$ is defined as $V_{TG}' = V_{TG} - V_{DP}$, $V_{ch}$ can be expressed as $V_{ch} = V_{TG}' - \int_0^{V_{TG}'} C'/C_{Y2O3} dV_{TG}'$.[5] Thus, the experimentally estimated $C_Q$ can be compared to the theoretical $C_Q$ ($=2e^2E_F/\pi(v_F\hbar)^2$), where $v_F$ is the Fermi velocity ($1\times10^8$ cm/s) and $\hbar$ is the Planck's constant. The experimentally estimated $C_Q$ is in good agreement with the theoretical $C_Q$ for $|E_F| > \sim0.15$ eV.

On the other hand, a large deviation from theory is evident near the DP, suggesting that more carriers exist than theoretically predicted. To elucidate whether these carriers are induced by the charged impurities or by the thermal excitation, the temperature dependence of the minimum $C_Q$ was measured. As shown in the inset of Fig. 3(b), almost no temperature dependence of the minimum $C_Q$ is observed. Thus, the residual carrier density $n^*$ near the DP is induced by charged impurities[3,4] and is calculated to be $3.6\times10^{11}$ cm$^{-2}$ using $E_F = \hbar v_F \sqrt{\pi n}$, as indicated by the arrow in Fig. 3(b). It should be noted that the minimum $C_Q$ value in the inset is different from that in main figure, since these data were obtained from a different sample.

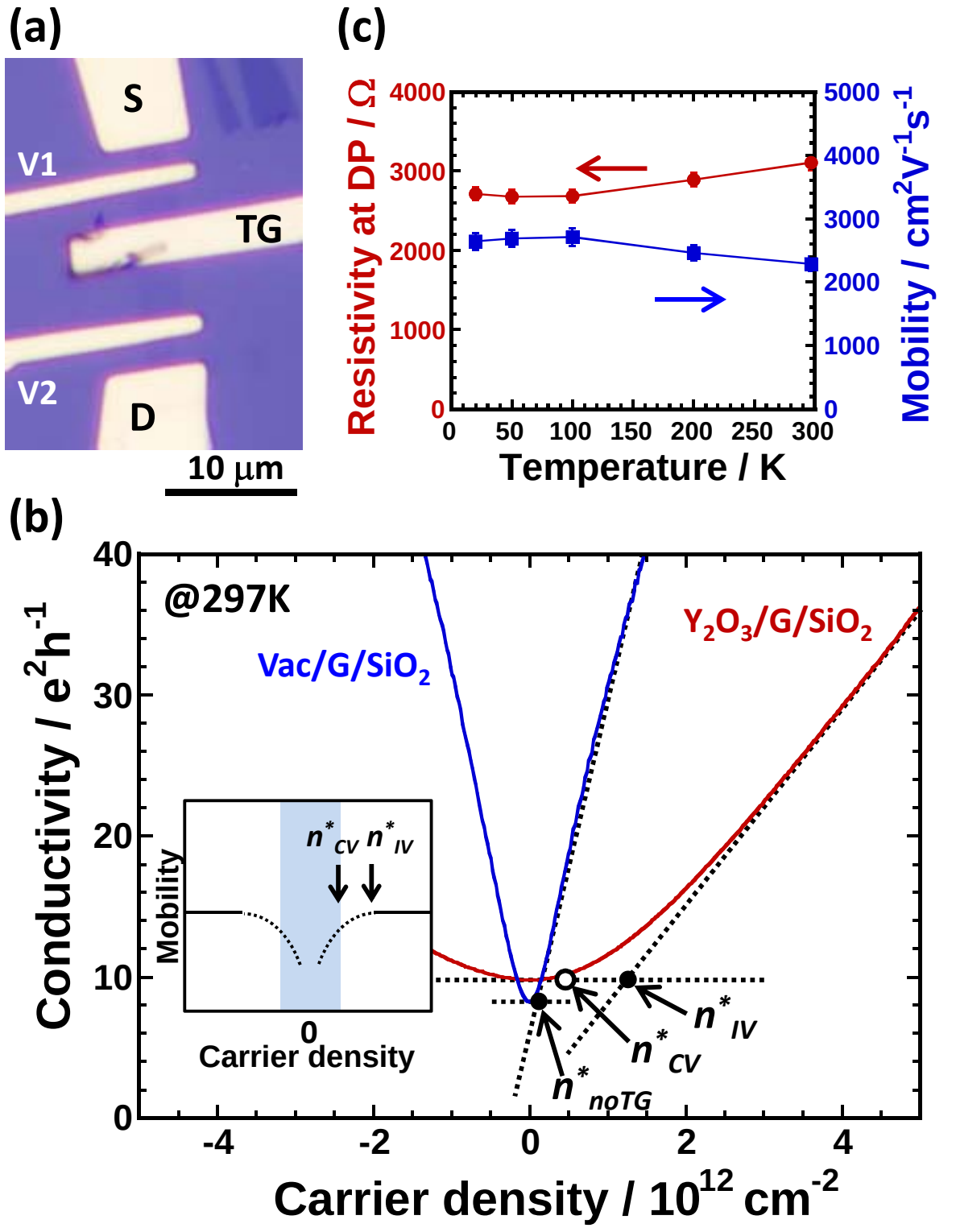

**FIG 4** (color online) (a) An optical micrograph of the four-probe dual-gated graphene FET device. (b) The $\sigma$ - $n$ relations for a graphene FET device with an $Y_2O_3$ top-gate insulator (red line) and without a top-gate insulator (blue line). The inset shows the schematic illustration of $\mu$ as a function of $n$. The hatched region is the electron-hole puddle region determined by the $C$-$V$ measurement. (c) The temperature dependence of resistivity at the DP and $\mu$ at n=$3\times10^{12}$ cm$^{-2}$ for the device in (a).

The residual carrier density $n^*$ near the DP is usually discussed from the $\sigma$ - $n$ relationship because $n^*$ is referred as the carrier density in the so-called puddle transport.[9] No comparison between $n^*_{I\text{-}V}$ and $n^*_{C\text{-}V}$, however, has been performed in the same graphene FET. In fact, it should be considered that the conductance is determined by not only the residual carriers but also the carrier mobility. Therefore, the four-probe dual-gated graphene FET was fabricated to remove the contribution of the contact resistance, as shown in Fig. 4(a). The $n^*_{I\text{-}V}$ was determined to be $1.25\times10^{12}$ cm$^{-2}$ from the intersect (solid circle) of the two dotted lines in Fig. 4(b), while $n^*_{C\text{-}V}$ for the same FET was determined to be $3.8\times10^{11}$ cm$^{-2}$ from the $C_Q$ - $E_F$ relation, which was obtained in the same manner for Fig. 3(b). The value of $n^*_{C\text{-}V}$ is also indicated in Fig. 4(b) as an open circle. The $n^*_{I\text{-}V}$ is larger than $n^*_{C\text{-}V}$. It should be noted that $n^*_{C\text{-}V}$ for seven different samples measured in this work lie in the range of $1.9\text{-}5.4\times10^{11}$ cm$^{-2}$ and are always smaller than $n^*_{I\text{-}V}$. In the analysis of the $I$-$V$ measurements, the simple two-band model[21], in which the same and constant mobility values for electron and hole are assumed for all carrier densities, was employed for explaining the conductivity ($\sigma$) - $n$ relation. On the other hand, $n^*_{C\text{-}V}$ is directly determined from the $C$-$V$ measurement, where the contribution of the scattering problem is excluded. Thus, the fact that $n^*_{I\text{-}V}$ is larger than $n^*_{C\text{-}V}$ suggests that the electron and hole mobilities in the puddle-edge region are lower than those in the linear $\sigma$ - $n$ region, as shown schematically in the inset of Fig. 4(b). Here, $n^*_{no_TG}$ in Fig. 4(b) indicates the residual carrier density for back-gated FET device determined by the $I$-$V$ measurement, which is smaller than $n^*_{C\text{-}V}$. It is evident that the deposition process of the $Y_2O_3$ top-gate insulator on graphene affects $n^*_{I\text{-}V}$. Therefore, the $C_Q$ measurement is necessary for estimating $n^*$ quantitatively and independently of the sample quality.

Finally, let us discuss the carrier scattering in the case of high-$k$ gate stack graphene FETs. There is a debate regarding whether high-$k$ dielectric materials on graphene improve the mobility due to a screening of the scattering potential from charged impurities on/in $SiO_2$ substrates.[22-24] Moreover, the remote phonon scattering due to the polar high-$k$ oxide with a low phonon energy may become the dominant scattering source rather than the Coulomb scattering by charged impurities.[25] In the remote phonon scattering, the mobility should be considerably suppressed and is strongly dependent on the temperature. To obtain more information on the scattering mechanism, the four-probe dual-gated FET, as shown in Fig. 4(a), was investigated. The $\sigma$ - $n$ relation in $Y_2O_3$ top-gated devices is linear at the high $n$ region as shown in Fig. 4(b). It should be noted that this $\sigma$ - $n$ curve was obtained by changing $V_{BG}$ because the top gate cannot electrostatically control the channel region between the top gate and the voltage probe. Figure 4(c) shows the resistivity at the DP and the electron mobility at $n = 3\times10^{12}$ cm$^{-2}$ in the linear $\sigma$ - $n$ region, as functions of the measurement temperature. Although the very weak temperature dependence of the resistivity at the DP is observed in the high temperature region (T > 200 K) possibly due to the remote phonon scattering from $Y_2O_3$ topgate or $SiO_2$ substrate,[26] it did not show the temperature dependence at low temperature. The very weak temperature dependence of the mobility also agrees with that reported previously for the graphene FET with the $HfO_2$ top gate,[27] since the main origin seems to be the Coulomb scattering due to the charged impurity on/in $Y_2O_3$ and $SiO_2$. These results indicate that the remote phonon scattering is not dominant but that the Coulomb scattering by charged impurities is still the main limiting factor. This is consistent with that for the minimum $C_Q$ (Fig. 3(b) inset). Moreover, the carrier mobility in the dual-gated graphene FETs is generally lower than those of typical back-gated graphene FETs without the top gate,[10] as is seen from the comparison of slopes in the $\sigma$ - $n$ relations in Fig. 4(b). This suggests that the improvement of the top gate stack formation on graphene may improve the mobility of dual-gated graphene FETs.

In summary, we performed *C-V* and *I-V* measurements on the same dual-gated graphene FET. Both the minimum $C_Q$ and the resistivity at the DP have shown nearly zero temperature dependence, indicating that carriers are induced by charged impurities. Moreover, it has been found experimentally that $n^*_{C-V}$ is often smaller than $n^*_{I-V}$. This fact has been discussed from the viewpoint that the carrier mobility in the edge of the puddle region is lower than that in linear conductivity region. Thus, it is concluded that the $C_Q$ measurement should be employed for estimating $n^*$ quantitatively. In addition, a relatively weak temperature dependence of the mobility and the linear $\sigma$ - $n$ relation suggest that the dominant scattering source in the high-*k* top-gated FETs used in this study is the Coulomb scattering by charged impurities.

**Acknowledgements**

Kish graphite used in this study was kindly provided by Covalent Materials Co. This work was partly supported by a Grant-in-Aid for Scientific Research from The Ministry of Education, Culture, Sports, Science and Technology, Japan.